\documentclass[sigplan,nonacm]{acmart}
\usepackage[utf8]{inputenc}
\usepackage{href-ul}
\usepackage{natbib}
\usepackage{debate}
\usepackage[noframes]{ffcode}
\usepackage{tabularx}
\usepackage{booktabs}
\usepackage{href-ul}
\usepackage{to-be-determined}
\usepackage{float}
\usepackage[toc,page]{appendix}
\usepackage[capitalize]{cleveref}
\usepackage{paralist}
\usepackage{colortbl}

\title{Embracing Objects Over Statics: An Analysis of Method Preferences in Open Source Java Frameworks}

\author{Vladimir Zakharov}
\orcid{0009-0007-2420-5675}
\email{volodya.lombrozo@gmail.com}
\affiliation{\institution{Huawei}\country{Russia}\city{Moscow}}

\author{Yegor Bugayenko}
\orcid{0000-0001-6370-0678}
\email{yegor256@gmail.com}
\affiliation{\institution{Huawei}\country{Russia}\city{Moscow}}


\begin{document}

\begin{abstract}
In today's software development landscape, the extent to which Java applications utilize object-oriented programming paradigm remains a subject of interest. Although some researches point to the considerable overhead associated with object orientation, one might logically assume that modern Java applications would lean towards a procedural style to boost performance, favoring static over instance method calls. In order to validate this assumption, this study scrutinizes the runtime behavior of 28 open-source Java frameworks using the YourKit profiler. Contrary to expectations, our findings reveal a predominant use of instance methods and constructors over static methods. This suggests that developers still favor an object-oriented approach, despite its potential drawbacks.
\end{abstract}

\maketitle

\section{Introduction}\label{sec:introduction}

Java programs, as well as those in some other object-oriented programming languages, must be composed of classes~\citep{rentsch1982object}. These classes, on one hand, serve as template definitions for the methods and variables in objects~\citep{stroustrup1988object}. On the other hand, classes may be used as namespaces for globally accessible procedures known as static methods. Because of this duality, Java may be considered as a hybrid programming language that mixes objects and procedures.

The object-oriented design paradigm, as it was originally proposed in Simula~\citep{dahl1970simula} and Smalltalk~\citep{kay1996early}, implies that objects communicate through dynamically dispatched method calls, enabling declarative programming with abstraction and polymorphism~\citep{bacon1996fast,pande1996data}. Thus, even though static methods are used in some object-oriented design patterns, such as Singleton~\citep{gamma1995elements} and Static Factory Method~\citep{bloch2008effective}, they must be attributed to imperative programming features.

One of the key differences between imperative and declarative programs is that the former are faster to run but more difficult to maintain than the latter~\citep{chambers1991making}.
Dynamic dispatch along with object allocations are the two primary sources of performance inefficiencies~\citep{driesen1996direct,budimlic1999cost}.
Because of this, it is expected that programmers tend to use fast static methods over slow polymorphic objects, especially in open source frameworks and libraries. However, this intuition has not been confirmed by any systematic study as of yet.

In order to fill this gap, we created and then, with the help of YourKit profiler, analyzed the runtime behavior of nine Java programs. We discovered that 28 open source frameworks that were used by our programs tend to favor the object-oriented approach over the object-less procedural one, despite the associated challenges: static methods comprise merely 12\% of all method invocations, 73\% are instance methods, and the rest are constructors.

We believe these findings underscore the continued significance of the object-oriented paradigm for most applications and libraries. Such results may suggest that program maintainability is often prioritized over computational efficiency. This underlines the need for future research to mitigate potential downsides of the object-oriented paradigm and to introduce more efficient optimizations in this area.

The rest of the paper has the following structure: 
\Cref{sec:example} presents a motivating example to demonstrate performance effect of object-orientation in Java. \Cref{sec:related} offers a review of existing studies that assess the impact of the object-oriented approach. In \cref{sec:method}, we detail the method used to collect statistics from Java applications. \cref{sec:results} presents the outcomes derived from the method described in \cref{sec:method}. \cref{sec:discussion} provides an overview of these results and sheds light on future research. Lastly, \cref{sec:conclusion} summarizes the findings of this study.

\section{Motivating Example}\label{sec:example}

A Java program in \cref{fig:procedural-simple-sample} uses only static methods and instantiates no objects. It runs eight times faster than its functionality equivalent counterpart in \cref{fig:oop-simple-sample}, which uses Decorator Pattern~\citep{gamma1995design}. The experiment was done with Java OpenJDK Zulu17.38+21-CA (build 17.0.5+8-LTS) compiler, with JIT enabled. In every program, there is a loop of 40 million iterations, which is a large enough number to let JIT ``warm up'' and compile Bytecode into machine code. We tested both snippets with and without garbage collection (by disabling it): the execution time remained approximately the same. 

\begin{figure}
\begin{ffcode}
public class Procedural {
  static int foo(int u) { return u; }
  static int bar(int u) { return u / 2; }
  static void main(String... args) {
    int sum = 0;
    for (int i = 0; i < 40_000_000L; ++i) 
      { sum += bar(foo(i)); } } }
\end{ffcode}
\caption{Java program that uses procedural approach, focuses on simple computations using static methods without utilizing of any objects.}    
\label{fig:procedural-simple-sample}
\end{figure}

\begin{figure}
\begin{ffcode}
class ObjectOriented {
  interface Type { int total();}
  static class Foo implements Type {
    private int i; Foo(int i) { i = i; }
    @Override public int total() { return i; } }
  static class Bar implements Type {
    private Type t; Bar(Type t) { t = t; }
    @Override public int total() 
      { return t.total() / 2; } }
  static void main(String... args) {
    int sum = 0;
    for (int i = 0; i < 40_000_000; ++i) { 
      sum += new Bar(new Foo(i)).total(); } } }    
\end{ffcode}
\caption{Java program that uses object-oriented approach, performing the same computations as the program of \cref{fig:procedural-simple-sample}, but emphasizing high objects usage through the Decorator Pattern.}
\label{fig:oop-simple-sample}
\end{figure}

\Cref{fig:procedural-bytecode} shows Bytecode of the procedural program of \cref{fig:procedural-simple-sample}, while \cref{fig:oop-bytecode} shows Bytecode of the object-oriented program of \cref{fig:oop-simple-sample}. Obviously, the cumulative performance cost of five |invokespecial|, one |invokevirtual|, and one |invokeinterface| together is higher than the cost of two |invokestatic|~\citep{ishizaki2000study, virtual-calls-lee2000reducing, virtual-calls-sundaresan2000practical}.

\begin{figure}
\begin{ffcode}
public class Procedural
 static int foo(int);
  0: iload_0
  1: ireturn
 static int bar(int);
  0: iload_0
  1: iconst_2
  2: idiv
  3: ireturn
 static void main(java.lang.String...);
  15: invokestatic  #9  // Method foo:(I)I
  18: invokestatic  #15 // Method bar:(I)I
\end{ffcode}
\caption{Bytecode for the ``procedural'' Java program of \cref{fig:procedural-simple-sample}, where some less relevant lines were removed for the sake of brevity.}
\label{fig:procedural-bytecode}
\end{figure}

\begin{figure}
\begin{ffcode}
class ObjectOriented
 static void main(java.lang.String...);
  10: iload_1
  11: new #8 // class Bar
  14: dup
  15: new #10 // class Foo
  18: dup
  19: iload_2
  20: invokespecial #12 // Foo."<init>":(I)V
  23: invokespecial #15 // Bar."<init>":(LType;)V
  26: invokevirtual #18 // Bar.total:()I
class Foo implements Type
 private int i;
 public int total();
  0: aload_0
  1: getfield #7 // Field i:I
  4: ireturn
class Bar implements Type {
 private Type t;
 public int total();
  0: aload_0
  1: getfield #7 // Field t:LType;
  4: invokeinterface #13, 1 // Type.total:()I
  9: iconst_2
  10: idiv
  11: ireturn
\end{ffcode}
\caption{Bytecode for the object-oriented Java program of \cref{fig:oop-simple-sample}, where some less relevant lines and constructors were removed for the sake of brevity.}
\label{fig:oop-bytecode}
\end{figure}

There are two sources of performance degradation in object-oriented programs: virtual calls (polymorphism) and object allocations. Virtual methods calls are implemented in Java through |invokevirtual|, |invokeinterface|, and |invokedynamic| opcodes, which are more expensive than |invokestatic| that is used for static method calls (without the use of virtual tables)~\citep{ishizaki2000study}. Object allocations are implemented in Java by the |new| opcode, which is not required in case of a procedural approach and the use of static methods: no objects are allocated at all. Even without polymorphism and dynamic dispatching, object and static method calls  would have different performance, because object allocations would still be required.

The intent of this example was to demonstrate that the usage of objects in Java, especially with polymorphism and dynamic dispatch, has certain performance costs associated with it~\citep{virtual-calls-lee2000reducing}. Despite available mechanisms of optimization, such as object inlining~\citep{object-inlining-lhotak2002run}, both Java compiler and Java Virtual Machine can't make polymorphic objects as fast as static methods~\citep{virtual-calls-sundaresan2000practical}. This fact may lead to a conclusion that programmers should avoid object-oriented programming (especially polymorphic objects) and design their code in procedural way. The aim of the research was to find out whether programmers indeed use static methods more often than object methods.

\section{Related Work}\label{sec:related}

Since the very birth of object-oriented programming, it was obvious that abstraction~\citep{bacon1996fast} and polymorphism~\citep{pande1996data} would lead to performance degradation~\citep{chambers1991making}. A number of studies validated this intuition, demonstrating that there is a substantial cost associated with a full object-oriented design~\citep{budimlic1999cost,chatzigeorgiou2002evaluating,chantarasathaporn2007object}, which is ``easier to write but slower to run''~\citep{chambers1991making}.

Since then, multiple optimization methods were introduced, attempting to reduce the negative performance effect of object-orientation, including inline caching~\citep{deutsch1984efficient}, 
object specialization~\citep{andersen2004declarative,vasileva2023object}, 
object inlining~\citep{dolby1997automatic,budimlic1997optimizing,dolby2000automatic,wimmer2007automatic},
devirtualization~\citep{ishizaki2000study},
semantic inlining~\citep{wu1998improving},
method specialization~\citep{schultz2003automatic}, 
object combining~\citep{veldema2002object},
object fusing~\citep{wimmer2010automatic}, 
elimination of lambdas~\citep{moller2020eliminating}, 
and
value fields inlining~\citep{bruno2021compiler}. 
However, even though some of these optimizations were implemented in HotSpot~\citep{hotspot-optimizations-paleczny2001java}, Java object methods are still slower than static methods, as it was demonstrated in \cref{sec:example}. It is possible that they will never be equally fast due to fundamental differences between object-oriented and procedural programming paradigms~\citep{holzle1991optimizing}.

There is a substantial amount of research into object-oriented software performance~\citep{maplesden2015performance}, in particular related to studying of runtime bloat~\citep{xu2010software}, memory leaks~\citep{xu2013precise}, and object churn~\citep{dufour2007blended}. All of these studies confirm from different angles that object-oriented programming techniques are performance wise more expensive than their procedural counterparts.

However, the question of whether Java programmers prefer the procedural programming style with static methods because of its better performance, despite the object-oriented style with polymorphic objects promising better maintainability~\citep{west2004object,eo1}, remains unanswered. To the best of our knowledge, no systematic study has been conducted so far that analyzes how often object methods and constructors are called in Java programs compared to static methods.

\section{Method}\label{sec:method}

The intent of this research was to compare the frequency of object instantiations and their methods calls with the frequency of static method calls, in runtime within open-source Java software. Our research methodology, detailed in this Section, was designed in several steps. The specific results obtained in each step are summarized in \cref{sec:results}.

Step~1: We selected a few open-source Java frameworks that satisfied all of the following criteria:
\begin{inparaenum}[a)]
\item more than 300 GitHub stars,
\item more than 150 GitHub forks,
\item more than 20K non-commenting lines of code,
and
\item created earlier than five years ago.
\end{inparaenum}
Obviously, the selection criteria was not ideal, for example because, as it was demonstrated by \citet{munaiah2017curating}, the number of stars may not always be a proxy for project quality or relevance. That's why we added one more criteria to the list: each framework must be mentioned by the curated list of ``awesome'' Java frameworks, maintained in GitHub by the community: \href{https://github.com/akullpp/awesome-java}{akullpp/awesome-java}.

Step~2: We created a few Java programs that utilized selected frameworks. Some frameworks provided executable demo programs as examples---we opted to use them instead of writing our own programs. We opted not to use automated tests, which were available for some frameworks, primarily due to the following two reasons:
\begin{inparaenum}[1)]
\item it is hard to install a complete testing environment from the original project, especially for large projects with multiple layers of testing
and
\item the results may be biased, as certain tests might be intentionally repeated multiple times with varying values, thus leading to specific behavior patterns, with the same program paths being tested disproportionately more than others---a scenario we aimed to avoid, as suggested by \citet{horky2015and}.
\end{inparaenum}
We did not intent to maintain one-to-one relationship between programs and frameworks: some programs used more than one frameworks.

\label{text:application-types}
All programs were Java applications, packaged as JAR files and executable from the command line. These programs ran indefinitely in foreground mode and interacted through their APIs. For web applications, HTTP was used; for the Derby database, the JDBC API was employed; and for Apache Kafka, the communication was facilitated via the raw TCP protocol. We intentionally selected these types of programs for several reasons:
\begin{inparaenum}[a)]
\item they are capable of running continuously for extended periods, allowing YourKit to collect comprehensive performance statistics,
\item it is relatively easy to generate test traffic with the help of Apache JMeter, as it supports all the aforementioned protocols, and
and
\item these programs often come with extensive documentation, which accelerates the setup process for profiling.
\end{inparaenum}

Step~3: We compiled the programs developed with the frameworks and obtained pre-compiled binaries for the remaining programs directly from official sources. Then, we executed each program alongside YourKit Java Profiler\footnote{\url{https://www.yourkit.com/}}, a commercial Java profiling agent that instruments Bytecode and, in runtime, counts the number and duration of method invocations. We selected YourKit over other tools for several reasons:
\begin{inparaenum}[a)]
\item it employs Bytecode instrumentation, which provides exact method invocation counts (this contrasts with sampling techniques, like those used by the async-profiler, which only offer approximate results and lack the capability to profile the precise number of invocations);
\item it supports extensive features and various profiling modes, which were vital during our experiments as we sought to gather as much data as possible;
\item it offers the capability to export profiling results in CSV format, which was instrumental for our subsequent analyses;
\item it includes an API that enables integration with the scripts, thereby allowing the construction of profiling pipelines as code;
and
\item it is accessible for free, which is beneficial for the research purposes, given its open-source support.
\end{inparaenum}
In contrast, other tools such as VisualVM, JProfiler, Eclipse MAT, and async-profiler, to the best of our knowledge, either lack some of these key features  or are commercially licensed, which restricts their use in our research context.

Step~4: With the help of Apache JMeter\footnote{\url{https://jmeter.apache.org/}}, we imposed heavy traffic on each program for one minute (the time sufficient enough to ``warm up'' JIT and run garbage collector a few times). We collected the counts of all invoked methods, including the methods of programs and of all utilized frameworks. Our main goal was to evaluate the application under conditions resembling typical real-world loads.  We considered several alternatives, including ApacheBench, WAPT Pro, HP LoadRunner, Siege, and Locust, but ultimately selected JMeter for several reasons:
\begin{inparaenum}[a)]
\item its open-source nature allows for cost-free usage in our research;
\item it offers a variety of settings and plugins, enabling precise load configuration for different APIs, such as HTTP, TCP, and JDBC;
\item it features a robust command-line interface that facilitates integration into our profiling scripts;
and
\item it has extensive documentation and widespread industry adoption, simplifying its use.
\end{inparaenum}
Furthermore, there are some studies that also underscore better results provided by JMeter than by any other testing tools~\citep{ jmeter-abbas2017comparative}.

Step~5: We analyzed the list of all method calls provided by YourKit and extracted from it the list of all mentioned Java packages. Then, we analyzed the list of dependencies (JAR libraries) of all programs and matched package names with dependencies and their versions. Thus, we discovered which particular methods from which frameworks were invoked during the profiling session of the previous step.

Step~6: However, the data provided by YourKit did not include metadata about the methods being called: whether they were instance methods, static methods, or constructors. To add this information to the dataset, we parsed the Java source code of all classes in all frameworks and created a list of unique method names, along with their package names and modifiers. Then, we matched the data provided by YourKit with the modifiers of all methods collected during the Java parsing.

Step~7: We generated aggregated statistics for each framework and visually analyzed the results. Specifically, we identified the most frequently used method types and visualized these using histograms and tables.

\section{Results}\label{sec:results}

We conducted the experiments according to the plan described in \cref{sec:method}, on a machine running macOS~12.5, equipped with an Intel(R) Core(TM) i7-9750H CPU 2.60GHz and 16GB RAM. We utilized Java OpenJDK Zulu17.38+21-CA (build 17.0.5+8-LTS). The profiling tool employed was YourKit Java Profiler 2022.9-b183. Apache JMeter 5.5 was utilized to add artificial load to the profiling applications. All components were executed on the same machine.

Step~1: We selected nine Java frameworks that met our criteria:
\begin{inparaenum}[1)]
\item Spring MVC\footnote{\url{https://www.spring.io/}},
\item Struts\footnote{\url{https://struts.apache.org/}},
\item Takes\footnote{\url{https://www.takes.org/}},
\item Derby Database\footnote{\url{https://db.apache.org/derby/}},
\item Dropwizard\footnote{\url{https://www.dropwizard.io/}},
\item Kafka Message Broker\footnote{\url{https://kafka.apache.org/}},
\item Micronaut\footnote{\url{https://micronaut.io/}},
\item Tomcat\footnote{\url{https://tomcat.apache.org/}}, 
and
\item Vert.x\footnote{\url{https://vertx.io/}}.
\end{inparaenum}

Step~2: We created nine simple programs (web applications) using the mentioned frameworks. In three frameworks (Derby Database, Kafka Message Broker, and Tomcat), we found web apps already prepared for execution. The six programs we designed implemented a basic web controller that accepted HTTP requests with a body containing a JSON object, parsed the request, and constructed an HTTP response with another JSON object in the body.

Step~3: We compiled all programs using OpenJDK Java compiler and executed them with YourKit profiler agent, which instrumented Bytecode for subsequent profiling using the |-agentpath| option. Additionally, we incorporated the |_instrument_all_methods| option to the agent to ensure all methods were instrumented without exclusions.

Step~4: We subjected the applications to intensive loads using Apache JMeter test cases. Each test case consisted of a consistently repeated HTTP, JDBC and TCP requests. The request frequency ranged between 20--50 requests/sec. Through this method, JMeter simulated an application's runtime in basic scenarios.

\begin{figure*}
\includegraphics[width=\textwidth]{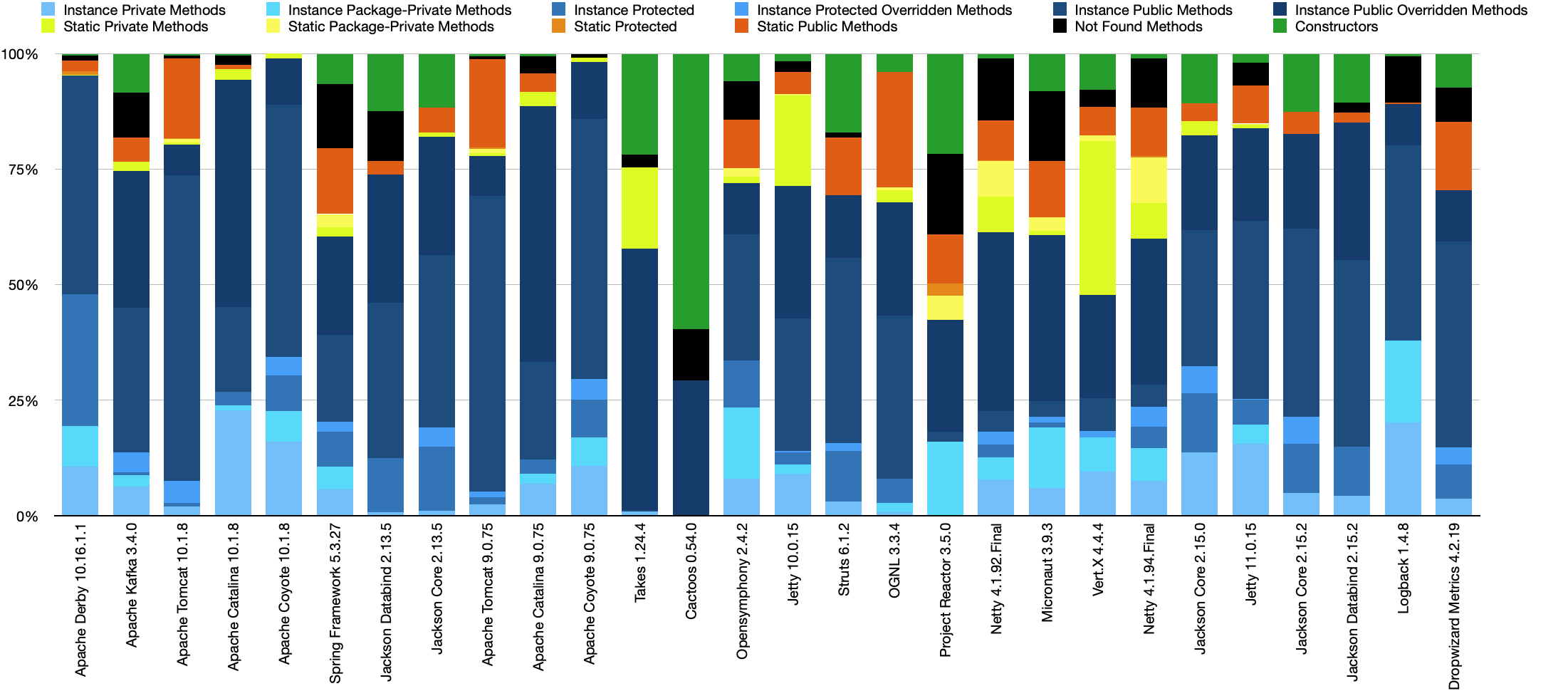}
\caption{Histogram of methods distribution: percentage of constructors, static methods, and instance methods across 28 analyzed frameworks. Blue indicates various types of instance methods, yellow and orange represent static methods, green denotes constructors, and black indicates methods that are generated at runtime and for which the type could not be determined.}
\label{fig:libraries-statistics}
\end{figure*}

Step~5: We exported all profiling results to CSV files, listing methods with their respective invocation counts and time spent on each. Typically, each CSV file cataloged 2,000--4,000 methods. Leveraging the names of Java packages, we identified the frameworks with more than 20,000 invocations (as represented on the horizontal axis of \cref{fig:libraries-statistics}). It is an empirically received number that indicates the considered framework has a high usage rate. Additionally, we verified that these invocations were approximately uniformly distributed among methods. This was done to exclude frameworks where only a few methods were invoked frequently.

Step~6: We downloaded the source code for each and merged the data from the CSV files with the source code by pinpointing methods' meta-information and modifiers.

Step~7: We generated the resultant CSV file, which displayed a list of frameworks with an aggregated count of instance methods (including private, package-private, protected, protected overridden, public, and public overridden instance methods), static methods (encompassing private, package-private, protected, and public static methods), and constructors, as illustrated by \cref{fig:libraries-statistics} and in \cref{tab:method-usages-per-libraries} (we grouped all types of static methods and all types of instance methods into corresponding columns).

We also analyzed how often instance methods, static methods, and constructors of different frameworks are called by different programs. \Cref{tab:jakson-core} demonstrates the usage of Jackson Core Library by three programs. The data shows that the distribution is very similar for all three programs. 

\begin{table*}
\caption{A summary of profiled programs and identified libraries used in the analysis: the first column lists program names, the second column describes the origins of programs. The third column enumerates the libraries used by the programs, as they were identified during the analysis.}
\label{table:programs_libraries}
\begin{tabularx}{\textwidth}{lXX}
\toprule
Program Code & Description & Libraries Used \\
\midrule
azure-deer & The raw binary for `Derby DB 10.16.1.1' was downloaded from \href{https://db.apache.org/derby/derby_downloads.html}{derby} without any user code and changes. & Apache Derby 10.16.1.1 \\
yellow-koala & The raw binary for `Apache Kafka 3.4.0' was downloaded from \href{https://kafka.apache.org/downloads}{kafka} without any user code and changes. & Apache Kafka 3.4.0 \\
crimson-tortoise & The raw binary for `Apache Tomcat 10.1.8' was downloaded from \href{https://tomcat.apache.org/download-10.cgi}{tomcat}. This version is unchanged and does not include any user code. & Apache Tomcat 10.1.8, Apache Coyote 10.1.8, Apache Catalina 10.1.8 \\
emerald-squirrel & Manually written application based on the Spring MVC and Spring Boot frameworks, Version 3.2.0. & Spring Framework 5.3.27, Jackson Databind 2.13.5, Jackson Core 2.13.5, Apache Tomcat 9.0.75, Apache Catalina 9.0.75, Apache Coyote 9.0.75 \\
tawny-tapir & Manually written application based on the Takes Framework, Version 1.24.4. & Takes 1.24.4, Cactoos 0.54.0 \\
sapphire-sheep & Manually written application based on the Struts 2 Framework, Version 6.3.0.2. & Opensymphony 2.4.2, Jetty 10.0.15, Struts 6.1.2, OGNL 3.3.4 \\
mint-manatee & Manually written application based on the Micronaut Framework, Version 3.10.4. & Project Reactor 3.5.0, Netty 4.1.92.Final, Micronaut 3.9.3 \\
vanilla-vole & Manually written application based on the VertX Framework, Version 4.4.4. & Vert.X 4.4.4, Netty 4.1.94.Final, Jackson Core 2.15.0 \\
black-donkey & Manually written application based on the Dropwizard Framework, Version 4.0.7. & Jetty 11.0.15, Jackson Core 2.15.2, Jackson Databind 2.15.2, Logback 1.4.8, Dropwizard Metrics 4.2.19 \\
\bottomrule
\end{tabularx}
\end{table*}

\begin{table*}
\caption{Percentage of different method types used within libraries.}
\label{tab:method-usages-per-libraries}
\begin{tabularx}{\textwidth}{X>{\ttfamily\raggedleft}p{5em}>{\ttfamily\raggedleft}p{5em}>{\ttfamily\raggedleft}p{5em}>{\ttfamily\raggedleft\arraybackslash}p{5em}}
\toprule
Library 
    & {\rmfamily Static method calls (\%)} 
    & {\rmfamily Instance method calls (\%)} 
    & {\rmfamily Constructor calls (\%)} 
    & {\rmfamily Other calls (\%)} \\
\midrule
Apache Derby 10.16.1.1 & 3.22 & 95.27 & 0.40 & 1.11 \\
Apache Kafka 3.4.0 & 7.36 & 74.50 & 8.43 & 9.70 \\
Apache Tomcat 10.1.8 & 18.65 & 80.32 & 0.51 & 0.51 \\
Apache Catalina 10.1.8 & 3.26 & 94.34 & 0.40 & 2.00 \\
Apache Coyote 10.1.8 & 1.13 & 98.86 & 0.00 & 0.02 \\
Spring Framework 5.3.27 & 19.10 & 60.45 & 6.66 & 13.79 \\
Jackson Databind 2.13.5 & 2.92 & 73.79 & 12.42 & 10.86 \\
Jackson Core 2.13.5 & 6.38 & 81.91 & 11.70 & 0.00 \\
Apache Tomcat 9.0.75 & 20.87 & 77.89 & 0.64 & 0.59 \\
Apache Catalina 9.0.75 & 7.18 & 88.56 & 0.67 & 3.59 \\
Apache Coyote 9.0.75 & 1.05 & 98.09 & 0.00 & 0.86 \\
Takes 1.24.4 & 17.65 & 57.72 & 21.90 & 2.74 \\
Cactoos 0.54.0 & 0.00 & 29.25 & 59.67 & 11.08 \\
Opensymphony 2.4.2 & 13.69 & 71.98 & 6.00 & 8.33 \\
Jetty 10.0.15 & 24.74 & 71.30 & 1.72 & 2.24 \\
Struts 6.1.2 & 12.54 & 69.34 & 17.08 & 1.04 \\
OGNL 3.3.4 & 28.14 & 67.83 & 4.03 & 0.00 \\
Project Reactor 3.5.0 & 18.41 & 42.42 & 21.66 & 17.51 \\
Netty 4.1.92.Final & 24.18 & 61.32 & 1.15 & 13.35 \\
Micronaut 3.9.3 & 16.00 & 60.76 & 8.22 & 15.02 \\
Vert.X 4.4.4 & 40.66 & 47.79 & 7.80 & 3.74 \\
Netty 4.1.94.Final & 28.43 & 59.87 & 1.03 & 10.67 \\
Jackson Core 2.15.0 & 6.86 & 82.35 & 10.78 & 0.00 \\
Jetty 11.0.15 & 9.28 & 83.77 & 1.92 & 5.02 \\
Jackson Core 2.15.2 & 4.85 & 82.52 & 12.62 & 0.00 \\
Jackson Databind 2.15.2 & 2.13 & 85.11 & 10.64 & 2.13 \\
Logback 1.4.8 & 0.33 & 89.09 & 0.66 & 9.91 \\
Dropwizard Metrics 4.2.19 & 14.79 & 70.41 & 7.40 & 7.40 \\
\bottomrule
\end{tabularx}
\end{table*}

\begin{table}
\caption{The percentage distribution of method usages, including instance methods, static methods, and constructors from the Jackson Core library, remains consistent across three different programs.}
\label{tab:jakson-core}
\begin{tabularx}{\columnwidth}{lX>{\ttfamily\raggedleft}p{5em}>{\ttfamily\raggedleft}p{4em}>{\ttfamily\raggedleft\arraybackslash}p{4em}}
\toprule
Version 
    & Program 
    & {\rmfamily Instance method calls (\%)} 
    & {\rmfamily Static methods calls (\%)}
    & {\rmfamily Construc\-tor calls (\%)}
    \\
\midrule
2.13.5 & emerald-squirrel & 81.9 & 6.4 & 11.7 \\
2.15.0 & vanilla-vole & 82.3 & 6.9 & 10.8 \\
2.15.2 & black-donkey & 82.5 & 4.9 & 12.6 \\
\bottomrule
\end{tabularx}
\end{table}

\section{Discussion}\label{sec:discussion}

\textbf{To what extent do open source Java frameworks employ the object paradigm?}
Our results indicate that most frameworks we studied lean towards the object-oriented approach, emphasizing object creation and instance method usage over procedural-style static methods. As depicted in \cref{fig:libraries-statistics}, about 73\% of invocations are for instance methods covering all visibility levels. Even though dynamic dispatch, a characteristic of instance methods, can introduce performance overhead and hinder micro-optimizations, developers consistently opt for this approach. Object creations account for about 8\% of all invocations, further solidifying the dominance of the object-oriented paradigm. Frameworks like Cactoos and the Takes Framework show even higher constructor usages at 60\% and 22\%, indicating varying adoption levels of object-oriented principles among frameworks.

\textbf{Why static methods are not used more often than object methods, while their performance advantages are obvious?}
In stark contrast, static methods, which require no object instances, comprise merely 12.2\% of invocations. This suggests a subdued embrace of the procedural approach, even in mature frameworks traditionally focused on optimization. It is clear that frameworks prefer the object-oriented paradigm over procedural methods, despite known shortcomings. Also, certain frameworks appear more committed to object-oriented principles than others.

\textbf{How representative is the data collected from just nine programs?}
Our study's scope, limited to nine programs, poses a limitation. Despite analyzing 28 frameworks, a larger dataset is essential to derive comprehensive results and more definitive conclusions. It would also be interesting, in future research, to analyze frameworks that make a much larger usage of static (or instance) methods than the average.

\textbf{How trustworthy are the data collected by YourKit?}
Indeed, benchmarking Java applications is a complicated process, where a number of factor must be taken into account, as suggested by \citet{georges2007java-performance-evaluation, horky2015and}. In our experiment, we addressed most of them:
\begin{inparaenum}[1)]
\item \emph{Realistic Load}: We subjected the programs to realistic loads. For web programs, this process included:
handling raw HTTP requests and parsing JSON in request bodies, calling business-level classes, and generating accurate HTTP responses with corresponding JSON-formatted bodies. For some programs, the load involved generating HTML files for responses. In the case of database (Apache Derby) and queue (Apache Kafka) profiling, we assessed performance by executing different operations such as addition, creation, and deletion different database rows or messages.
\item \emph{Concurrency Simulation}: For some programs (not for all), we emulated parallel loads to create a concurrent environment, utilizing multiple threads to impose the load.
\item \emph{JIT Compilation and Optimization Consideration}: All programs were profiled for a minimum of 60 seconds to accommodate JIT compilation and optimizations. This duration was experimentally determined to ensure program stability post the initial warm-up phase.
\item \emph{Minimized Profiling Influence}: By running the profiling application in a separate process, we reduced its impact on performance measurements. However, it is important to note that despite these precautions, the inclusion of profiling agents introduced a certain level of overhead, which was insignificant for the current research.
\end{inparaenum}

\textbf{Would results be significantly different if a different list of frameworks would be chosen?}
A large set of frameworks would definitely make our results more representative, but we don't expect the balance between numbers to be changed significantly. The frameworks that we studied do not belong to a particular open source community: they are made, for example, by Apache, Google, FasterXML, VMWare, and individual developers. We do not expect a significantly different programming style to exist in some other programming groups.

\textbf{How much the interference of YourKit instrumentation affected the experiment results?}
Obviously, YourKit's Bytecode instrumentation influenced the overall performance of programs. However, since the performance was not studied but only the count of method invocations, the effect of instrumentation is insignificant for the research.

\textbf{Why profiling approach was chosen and whether it is the most adequate for this kind of experiment?}
Indeed, the profiling results depend on how a framework is used by a program. A different workload may yield different results. The aim of the study was to find out how often, on average, Java programs use objects. Instead of statically analyzing the source code, we chosen run runtime approach because it detects the frequency of objects usage in, for example, loops and recursive calls. This is what static analysis fails short to detect.

\section{Conclusion}\label{sec:conclusion}

In this study, we explored the adoption of object-oriented paradigms across 28 Java frameworks by examining their runtime behavior in nine synthetic Java programs. Our findings reveal a distinct preference for the object-oriented approach, with object method invocations accounting for 73\% of all method calls. These results could incentivize developers of compilers and optimizers to develop innovative techniques to enhance the performance of objects and virtual method calls in Java and other object-oriented languages.

\section{Data Availability}

All scripts that we created for the research, tool configurations, and results in CSV files, are consolidated in a single Zenodo artifact: \url{https://zenodo.org/records/11060930}.


\small
\bibliographystyle{plainnat}
\bibliography{main}

\end{document}